# Accelerating Surface Composition Characterization of Thin-Film Materials Libraries using Multi-Output Gaussian Process Regression

F. Thelen[1], F. Lourens[1], A. Ludwig[1,*]

[1]Chair for Materials Discovery and Interfaces, Institute for Materials, Faculty of Mechanical Engineering, Ruhr University Bochum, Universitätsstraße 150, 44801 Bochum, Germany

Underlined authors contributed equally.

*corresponding author: alfred.ludwig@rub.de

**Abstract**

Efficient characterization of surface compositions across high-dimensional materials spaces is critical for accelerating the discovery of surface-dominated functional materials. While X-ray photoelectron spectroscopy allows detailed surface composition investigation, it remains a time-intensive technique. In this work, it is demonstrated that Gaussian process regression can be used to accurately predict surface compositions from rapidly acquired volume composition data obtained by energy-dispersive X-ray spectroscopy, drastically reducing the number of required surface measurements. As a proof of principle, an exemplary system, the oxide Mg-Mn-Al-O, is synthesized as a composition-spread thin-film materials library and analyzed by high-throughput methods. We show that the surface composition of the entire library can be predicted with an accuracy of 96% with only 13 measurements, reducing the total measurement time by 277 h. This is a scalable and data-efficient solution for integrating surface analysis into materials discovery workflows.

## 1. Introduction

For the development of surface-dominated high-performance materials (e.g. for applications in batteries, catalysis or solar water splitting cells), the exploration of large multidimensional search spaces using combinatorial deposition of thin-film materials libraries and high-throughput characterization (screening) is considered as an appropriate approach [1]. The surface is the topmost region of a material (here thin films) where the properties differ significantly from the volume. Often, the first few atomic layers of the thin film are the primary region of interest, as they influence many of the material's interactions with its environment [2]. In many cases, the easy-to-measure volume composition of an investigated thin-film composition spread, together with screening for a functional property, is sufficient to identify compositional regions of interest. However, relying on volume composition as a proxy for surface composition is misleading in cases where surface and volume compositions are different due to surface segregation or surface reactions. Therefore, direct measurement of the surface composition is essential for a more accurate understanding, but achieving this in a high-throughput manner is a significant challenge.

Several methods are available to assess the surface composition of thin films. Low energy ion scattering (LEIS) measures the topmost atomic layer and has been used for high-throughput surface composition studies [3]. However, it cannot resolve surface compositions of elements which are too close in atomic mass, and the quantification is further complicated by hydrocarbon contaminants typically adsorbed at the surface.

An alternative to LEIS is X-ray photoelectron spectroscopy (XPS) [4–7]. The probing depth of XPS is about 10 nm, which can be further reduced with angle-resolved XPS [8]. XPS features a low detection limit (appr. 0.1-1 at.%), high number of detectable elements (all except H and He), very few cases of peak overlaps that cannot be resolved, and the possibility of chemical state analysis[5]. However, in comparison to well-established high-throughput methods such as energy dispersive X-ray spectroscopy (EDX) for quantifying volume composition or LEIS, XPS needs significantly longer measurement times which depend on factors including the number of constituent elements. Only recording wide scans is often inadequate due to low resolution. Thus, each constituent element requires the acquisition of at least one narrow scan in high resolution, resulting in an increase in measurement duration with an increasing number of elements. The minimum quality of the spectra is another factor: longer acquisition times generally result in higher quality spectra, i.e., higher signal-to-noise ratios. While EDX can measure a single volume composition in about 40 s, the XPS measurement of a single surface composition can take up to 1-2 hours. XPS screenings of materials libraries with hundreds of compositions are therefore impractical, so that the bottleneck in accelerating surface composition analysis, e.g. to identify surface segregation, is the XPS measurement time. Additionally, the subsequent quantification and data analysis further extends the required characterization time, as for XPS this is more complex than for EDX.

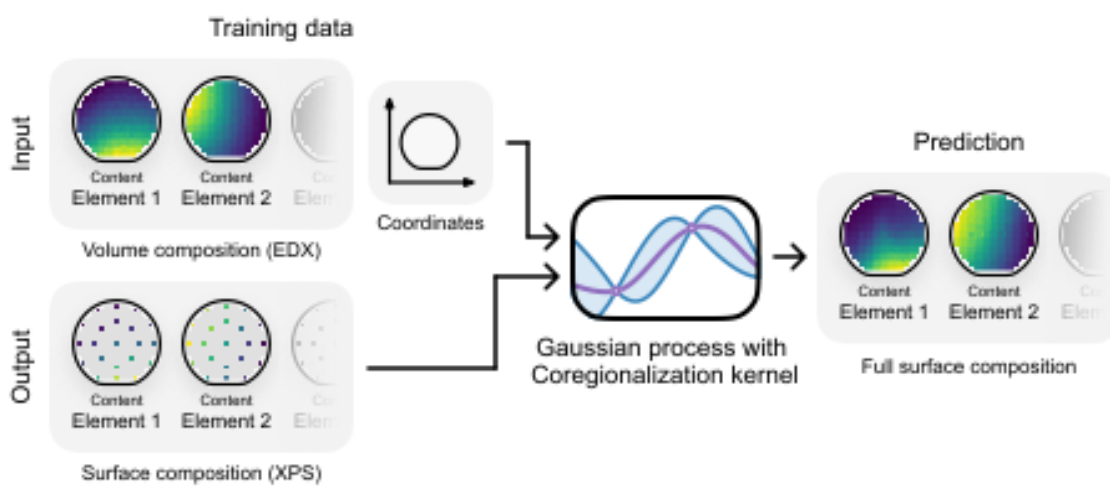

**Figure 1:** Structure of the Gaussian process for active learning: input (EDX data + coordinates) and output (measured XPS data) training data are connected to the Gaussian process and an illustration of the predicted XPS distribution over the entire materials library.

We propose to use active learning techniques in order to decrease the long measurement times of XPS: a surrogate machine learning model decides the measurement sequence by building and updating the model (e.g. a Gaussian process) during the procedure. Once the model's prediction is accurate enough, the process is terminated and the remaining areas are only predicted instead of measured, drastically reducing the total measurement time. Active learning algorithms are able to increase the measurement efficiency while still retaining an acceptable accuracy [9–11]. Ideally, active learning could be implemented in existing measurement devices if an application programming interface (API) is available, allowing to actively engage in the measurement procedure "on-the-fly". Even if an API is not available, we demonstrate that applying a Gaussian process still provides a substantial efficiency increase. By measuring an entire materials library once with XPS and simulating different numbers and arrangements of measurement areas and their effect on the Gaussian process predictions, we determine how many measurements are needed for maximum efficiency without sacrificing accuracy. For this, arrangements of measurement areas with maximum distances between them are compared to random arrangements.

Figure 1 shows the structure of the Gaussian process used. The 342 compositions of the entire materials library determined by EDX are used as input training data together with the x-y-coordinates of each measurement area. Since the model must be able to predict the content of the multiple elements in the multinary compound of interest, a multi-output Gaussian process with a Coregionalization kernel [12] is used. After hyperparameter optimization, the model is able to predict the surface composition of the entire materials library. Afterwards, the predictions are normalized to 100 at.% by dividing through the sum of all constituents' contents. This approach also allows to predict the surface oxygen content.

## 2. Results and Discussion

To demonstrate the application of Gaussian process predictions in the XPS setting, we revisit a materials library of the system Mg-Mn-Al-O, which was fabricated to investigate the formation of spinel solid solutions in slags derived from pyrometallurgical lithium-ion battery recycling in a preceding study. This library was chosen because, with 36 measurement areas already analyzed with XPS, a significant number of experiments was already available, providing a strong foundation for this study. While initial observations suggested a linear deviation between EDX and XPS data, more detailed investigations revealed non-linear surface segregation, making the selected multinary compound system an ideal case for testing the prediction capabilities of the Gaussian process. To allow reliable accuracy evaluation of the predictions, the effort was made to quantify the surface composition of the entire materials library comprising 342 measurement areas, resulting in a total XPS measurement duration of 12 complete days (288 h). Details about the fabrication and characterization of this materials library are provided in the experimental section. Figure 2 shows color-coded visualizations of the 342 measurement areas of the library for the EDX-volume-compositions and the XPS-surface-compositions. In accordance with the cathode positions in the combinatorial sputter chamber, each constituent element shows a compositional gradient along the diagonal of the library.

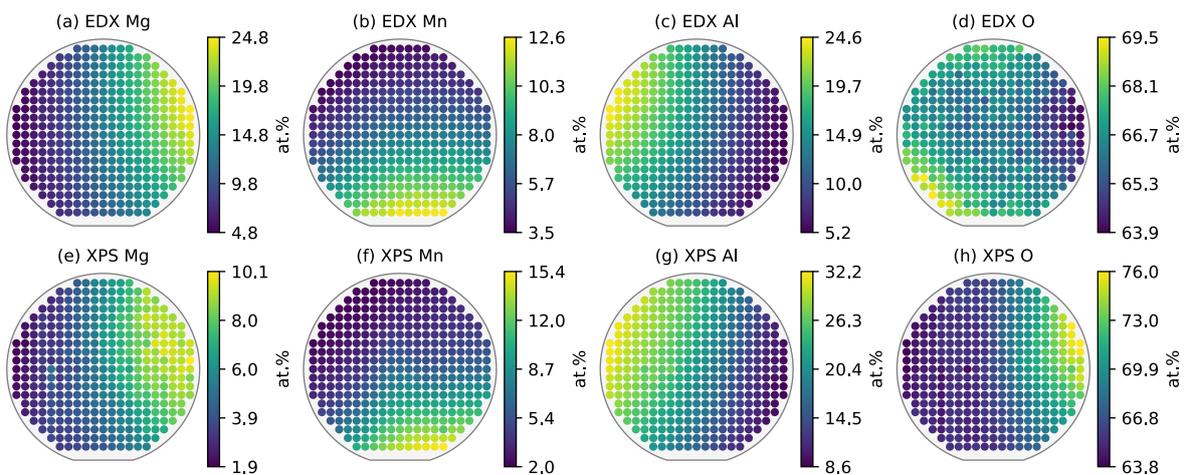

**Figure 2:** Color–coded visualization of the composition spreads of the thin film Mg-Mn-Al-O materials library for each constituent element: (a-d) Volume composition measured by EDX and (e-h) surface composition measured by XPS.

The compositional gradients of the deposited metals on the surface (Figures 2e-g) and in the volume of the thin film (Figures 2a-c) follow similar trends. The quantification of the oxygen content (and other elements with an atomic number lower than eight) is typically not reliable with EDX and the presence of oxygen in the protective $SiO_2$ layer of the Si substrate can further confound the measured signal. Therefore, oxygen quantified with EDX is excluded from further analysis.

Figure 3 visualizes the correlations between volume and surface compositions for all deposited metals and measurement areas. The results suggest a non-linear surface segregation with lower surface content of Mg and higher surface content of Al. In the case of Mn (Figure 3b), there are measurement areas with Mn-rich surfaces, Mn-poor surfaces, as well as areas where the surface and volume contents of Mn are equal. These findings are summarized in Figure 3d: a shift of the composition spread to the high-Al-low-Mg corner of the ternary composition space is observed.

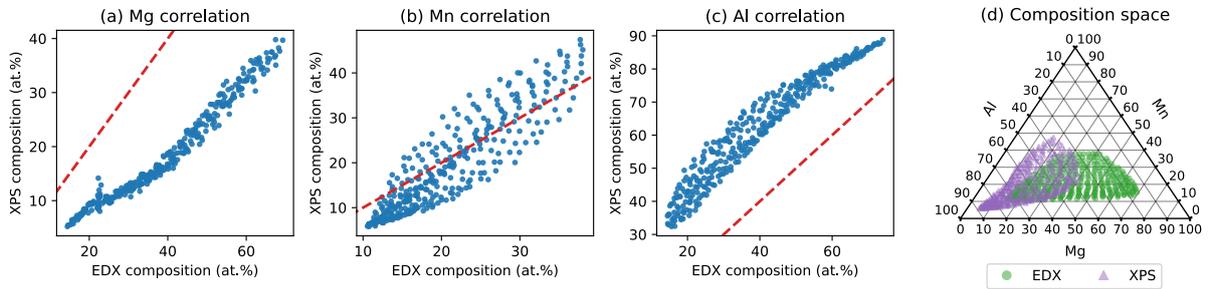

**Figure 3:** (a-c) Correlation between volume (EDX) and surface (XPS) compositions of Mg, Mn and Al. The red dashed lines indicate a perfect (equal) correlation. (d) EDX-XPS correlation visualized in the ternary composition space. All figures show normalized metallic compositions with excluded oxygen content.

Using this dataset, the number of measurement areas necessary to accurately predict the surface composition from the volume composition was investigated. However, the prediction accuracy additionally depends on the locations of the measurement areas selected as training data. Therefore, the prediction accuracy is investigated for different sets of randomly chosen XPS measurement areas and is compared to the accuracy achieved with evenly distributed ones. This comparison is based on the hypothesis that evenly distributed training data provides a superior prediction foundation than randomly selected areas due to improved coverage of the input space, reduced risk of extrapolation errors, and more balanced hyperparameter optimization. However, generating optimal measurement area arrangements on the predefined 342 area grid maximizing the distances between the areas is computationally expensive. Since the complexity is

$$O\left(\binom{342}{n} \cdot n^2\right)$$

with n being the number of areas, a greedy algorithm [13] is used to generate evenly distributed constellations, see Figure 4. In contrast to the brute force approach, this has a significantly lower complexity of $O(342 \cdot n)$.

The algorithm is initialized by selecting a starting area: here the center area of the library (Figure 4a). Then, the Euclidean distance between every other area and the already selected one is calculated. Then, the area with the maximum distance to the already chosen area(s) is selected next. An evenly distributed measurement area arrangement close to the global optimum is generated with feasible computational complexity by repeating this procedure until a desired number of areas is reached.

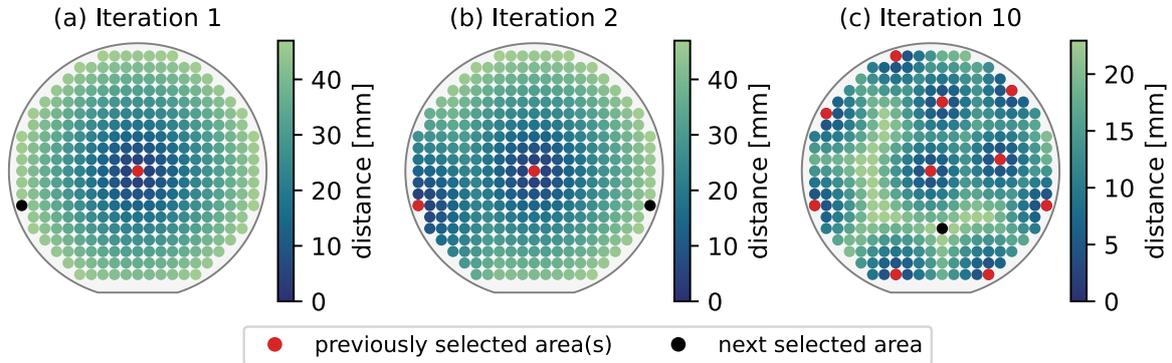

● previously selected area(s)   ● next selected area

**Figure 4:** Visualization of the greedy algorithm for selecting measurement areas evenly distributed across the measurement grid. The algorithm is initialized with the center area and the Euclidean distance to every other area is calculated (a). The area with the largest distance is selected next, and the distances to all other areas is calculated again (b). This is repeated until a specified number of areas was selected (c).

The prediction accuracy of the multi-output Gaussian process trained on up to 30 measurement areas (randomly or evenly distributed) is shown in Figure 5a. Four selected prediction results are plotted in the composition space in Figures 5b-e. The squared exponential kernel was used as the base kernel, and the accuracy was quantified using the coefficient of determination ($R^2$). At a low number of measured areas in the training dataset ($n_{areas}$), the model fails to accurately learn the underlying function. In these cases, the predictions stay close to the mean composition of the training data, which can be attributed to suboptimal hyperparameter optimization. This effect is evident in Figure 5b, where predictions cluster around the mean composition of $Mg_{17}Mn_{21}Al_{62}$ when trained on only four evenly distributed measurement areas. For an insufficient number of training points, the model tends to overestimate the length scale and the noise variance, leading to a constant surface composition prediction over the entire library. With six or more evenly distributed areas, the hyperparameter optimization improves significantly, resulting in an accuracy of 95% at $n_{areas} = 6$. As shown in Figure 5c, the predicted surface composition then closely resembles the measured distribution, with only minor deviations near the edges of the composition space covered in the materials library. With a higher number of training points (Figures 5d, e), the prediction accuracy increases, so that the prediction also closely resembles the edges of the composition space covered by the library. By measuring only 15 measurement areas instead of the entire materials library, the total measurement time is reduced to approximately 12.5 h, representing a 95% reduction in measurement duration.

Randomly distributed training areas can result in both an improved and a reduced predictive performance compared to evenly distributed training areas. However, since most of the models trained on randomly selected measurement areas show lower accuracy than the evenly distributed areas, evenly distributed areas are favorable. Since a random selection does not guarantee a wide distribution across the input space, hyperparameter optimization is more difficult, causing the model to require more training data until the length scale and variance are appropriately determined. Additionally, local clustering can cause the model to adapt to local variations rather than global trends.

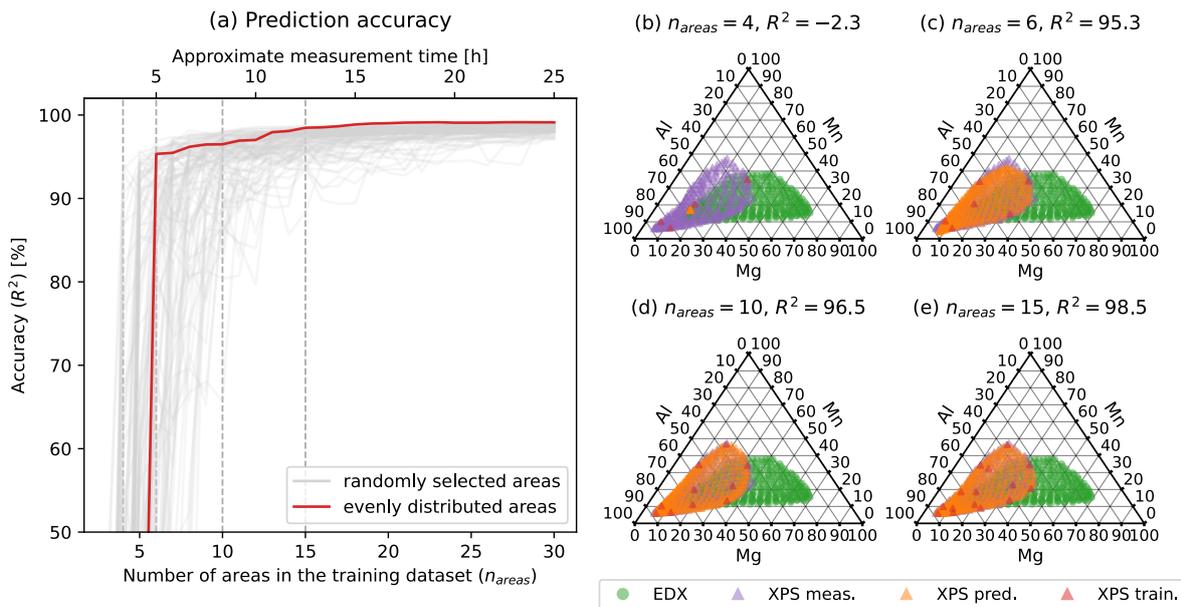

**Figure 5:** (a) Prediction accuracy over the number of measurement areas in the training dataset. 100 randomly generated measurement area arrangements on the 342-area grid are compared to evenly distributed areas generated by the greedy algorithm. The secondary x-axis allows to estimate the approximate XPS measurement time. (b-e) Surface composition predictions in case of 4, 6, 10 and 15 evenly distributed areas used as training data, alongside ground truth values.

The prediction performance of a Gaussian process can be strongly influenced by the choice of a kernel function. Therefore, the prediction accuracy achieved with the widely used Squared Exponential kernel is compared to that of the Rational Quadratic and Matérn kernels [14] in Figure 6a. In contrast to the Squared Exponential kernel, the Rational Quadratic and Matérn kernels include additional hyperparameters and therefore allow a more flexible fit. However, in this case, the increased flexibility appears to provide only a marginal improvement in prediction performance. For smaller numbers of training points, both the Rational Quadratic and Matérn kernels yield slightly higher prediction accuracies compared to the Squared Exponential kernel. Due to their higher flexibility, the kernels are able to reach 98% accuracy with two less areas, corresponding to a further measurement time reduction of approximately 1.5 h. Figure 6b shows the mean Euclidean distance of the prediction in dependance on the number of areas in the training dataset. Similar to the accuracy metric (Figure 6a), the Euclidean distance (Figure 6b) shows a gradual decrease of the prediction error with an increasing number of training areas. With 13 points in the training dataset, the prediction error falls well below 1 at.%.

The range of accuracies obtained from randomly selected measurement areas is indicated by shaded regions in Figure 6. Consistent with the results for evenly distributed measurement areas, the differences in performance among the kernels remain relatively small. This is likely attributable to the compositional trends across the materials library being broad and smooth, with limited local variation.

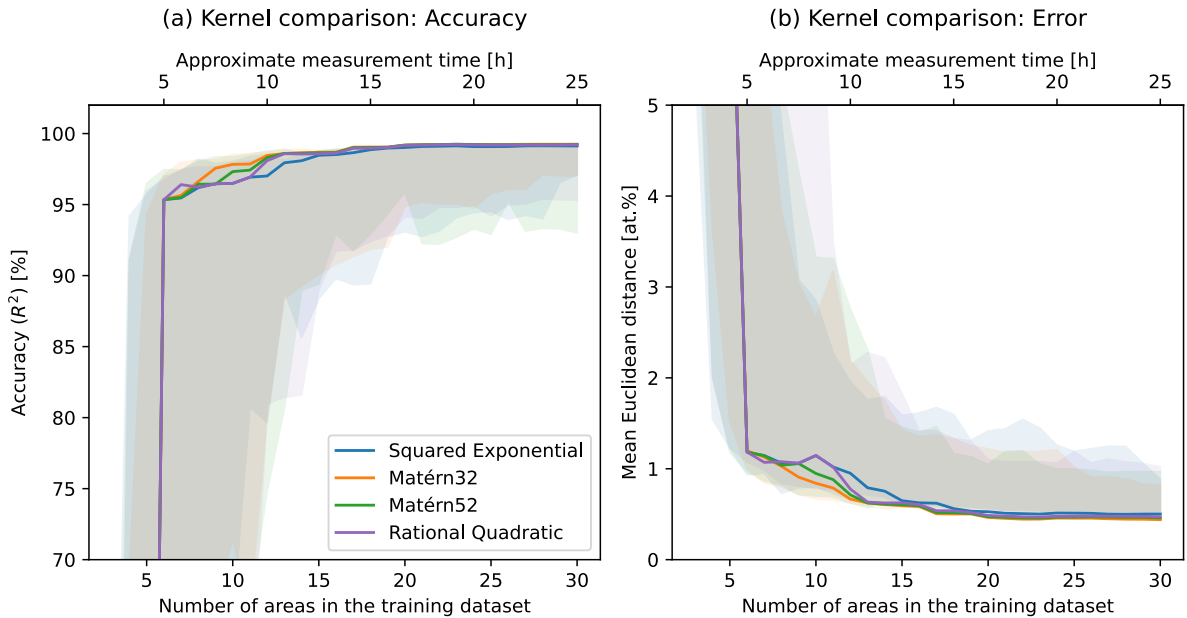

**Figure 6:** (a) Comparison of the prediction accuracy achieved with different kernels. Apart from the Squared Exponential kernel, the Rational Quadratic and two Matérn kernels were tested. (b) Prediction error measured with the Euclidean distance. Colored shadows denote the minimum and maximum metrics achieved with random selection of points. A small effect of the kernel choice on the accuracy is observed. The Matérn32 kernel shows a slightly better performance with fewer number of areas in the training dataset. From approximately 15 training points, the kernels behave identically except for minor deviations.

The results presented in this study demonstrate a significant efficiency improvement for the characterization of surface compositions. Only 13 XPS measurements are needed to achieve a near-perfect prediction using the Matérn32 kernel. The correlation of the ground truth and the prediction after training on the 13 evenly distributed areas is shown in Figure 7. The correlations indicate a strong generalization capability of the Gaussian process and also show that the oxygen content can be correctly predicted, even though oxygen is not part of the training data.

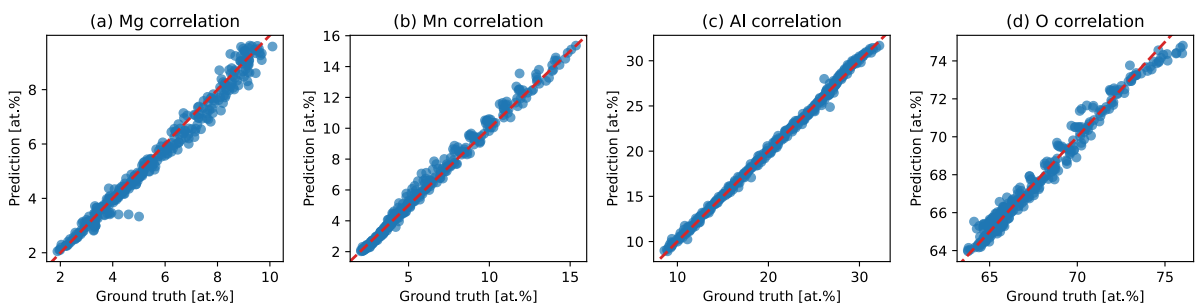

**Figure 7:** Correlation plots of ground truth (measured XPS compositions) and prediction when training the Gaussian process with the Matérn kernel on the 13 evenly distributed areas. Red lines indicate perfect correlation.

This performance is attributed to several contributing factors. Although the Mg-Mn-Al-O materials library shows a large nonlinear deviation between surface and volume composition, the overall spatial trends (across the coordinates of the library) in surface and volume compositions remain similar, enabling the model to learn a consistent transformation between the two. Moreover, the deviation of surface and volume content likely resides on a low-dimensional manifold, which simplifies the learning task for the model. The compositional gradients across the library are also continuous and smooth, creating favorable conditions for the kernels, and the limited local variation makes accurate interpolation possible with a small number of well-chosen inference points. Additionally, both volume and surface compositions feature low noise levels.

Since the investigated materials library shows a large nonlinear deviation of surface and volume composition, we expect that the presented method will be broadly applicable to other materials systems. For instance, libraries fabricated under non-reactive conditions and/or at room temperature are likely to show a smaller deviation of surface and volume composition, potentially requiring even fewer measurements to achieve comparable accuracy. However, it cannot be ruled out that libraries with more complex correlations also exist; yet systematically repeating this analysis across multiple materials libraries is not feasible due to the substantial measurement time required for full XPS characterization.

To ensure robust predictions across various systems, we propose a practical measurement strategy that involves measuring five additional areas beyond the minimal requirement, resulting in a total of 18 measurement areas. This modest increase enables the use of a leave-one-out validation strategy without significantly impacting measurement time. The 18 areas correspond to approximately 15 h of total measurement time, making the procedure compatible with an overnight measurement and offering a balance between efficiency and robustness, allowing further measurements to be selectively performed if necessary based on the validation outcomes.

## 3. Conclusions

This study demonstrates that the gap between high-throughput EDX and low throughput XPS can be closed by applying multi-output Gaussian process regression. By leveraging the correlation between volume and surface composition, we show that accurate predictions of the full materials library can be achieved with only a small subset of XPS measurements, which is of great significance for accelerating surface segregation studies. For the investigated Mg-Mn-Al-O materials library, a near-perfect prediction using only 13 evenly distributed measurement areas was achieved, corresponding to a 96% reduction in measurement time. The approach benefits from the smooth nature of the composition gradients, as well as the similarity between surface and volume trends. Further it is shown that the choice of Kernel function is of minor importance as different Kernel functions behave similarly in this scenario.

Due to the large deviation of surface and volume composition of the investigated materials library, it is expected that the findings can be generalized to other materials libraries fabricated in different materials systems or process conditions. Since more complex surface-volume correlations may exist, we proposed a successful leave-one-out strategy with five additional points to validate the prediction results while still allowing the measurement to be performed overnight.

The algorithm developed in this study provides a foundation for future active learning applications. As soon as an appropriate API becomes available for XPS instruments, this method can be adapted to actively guide measurements in real-time, dynamically optimizing the measurement sequence and further enhancing experimental efficiency.

## 4. Experimental section

The Mg-Mn-Al-O composition-spread type thin-film materials library was synthesized by reactive magnetron co-sputtering using three confocal cathodes arranged 120° apart from each other. The sputter targets were Mg (50.8 mm diameter, 99.95% purity, AJA International), Mn (101.6 mm diameter, 99.95% purity, Sindlhauser Materials), and Al (50.8 mm diameter, 100% purity, Kurt J. Lesker Company). The base pressure before deposition was $9.8 \cdot 10^{-6}$ Pa. A silicon wafer (100 mm diameter) with native oxide served as substrate and was heated to a deposition temperature of 300°C. The deposition was performed at a pressure of 2 Pa, with gas flows of 40 SCCM Ar (6.0, Praxair) and 2 SCCM O2 (6.0, Praxair). The sputter powers were 90 W (DC) for Mg, 120 W (RF) for Mn, and 70 W (DC) for Al. The deposition duration was 4 h.

A grid of 342 measurement areas (each 4.5 x 4.5 mm$^2$) distributed evenly across the whole library defined the measurement points for the EDX and XPS screenings. The EDX screening was carried out on a scanning electron microscope (SEM, JEOL 5800LV) with an EDX detector (Oxford INCA X-act). The EDX measurement conditions were: 20 kV acceleration voltage, 10 mm working distance, 600x magnification, and 60 s acquisition time per measurement area. The duration of the whole EDX screening including stage movement and other delays (e.g. software-related) was around 8 h. The XPS screening was carried out on an automated XPS (Kratos Axis Nova) with a monochromatic Al Kα X-ray source operating at 180 W (15 mA emission current, 12 kV anode voltage), and a delay-line detector with 20 eV pass energy. At each measurement area, high-resolution scans of 300 x 700 µm$^2$ analysis areas of the Mg 1s, Mn 2p, Al 2p, O 1s and C 1s regions were recorded. The latter was only used for charge correction (adventitious C-C at 284.8 eV), while the others were used for quantification, using Shirley backgrounds and the Kratos ESCApe software with its pre-defined relative sensitivity factors. The XPS acquisition time per area was around 30-35 min, comprising around 26-31 min for recording the high-resolution scans and 4 minutes for the obligatory sample height optimization and recording a wide scan. The duration of the whole XPS screening (including stage movement and other unavoidable delays in between individual scans) was around 12 complete days (288 h). Because the state of the X-ray filament changed and needed replacing, the XPS screening was divided into two subsequent sessions.

For predicting the surface compositions from the volume compositions, a multi-output Gaussian process with Coregionalization kernel implemented in GPflow [15] was used. A Gaussian process models a function $f(x)$ as a distribution over functions, such that any finite set of function values follows a multivariate normal distribution:

$$f(x) \sim GP(\mu(x), k(x, x'))$$

where $\mu(x)$ is the mean function and $k(x, x')$ is the covariance function [16]. The mean function is often assumed to be $\mu(x) = 0$, as the data can be standardized, and the Gaussian process is generally flexible enough to model the mean sufficiently well. The covariance function consists of a kernel function which defines the correlation of two random variables $x$ and $x'$ and therefore controls the function's shape. There are a number of different kernels, each with its own set of hyperparameters. A popular kernel choice is the squared exponential (SE) kernel:

$$k(x, x') = \sigma_f^2 \cdot \exp\left(\frac{-(x - x')^2}{2l^2}\right)$$

with $\sigma_f^2$ being the noise variance and $l$ the length scale parameter. Other kernels with a higher number of hyperparameters are the Matérn kernels and the rational quadratic kernel, which generally provide a more flexible fit [14].

In case of a multi-output regression problem, multiple standard Gaussian processes could be used to model each output independently. However, this would ignore the correlations between individual outputs. To address this, the covariance structure of the standard implementation can be extended by accounting both for the correlation between inputs and between outputs. This is achieved by introducing a second covariance matrix $K_f$, which models the relation between different outputs $l$ and $l'$, allowing information transfer across outputs and improving predictive performance. The multi-task covariance function can then be defined as [12]:

$$K_{coregion}\big((x,l),(x',l')\big) = K_f(l,l') \cdot k(x,x')$$

Since a Coregionalization kernel can account for the correlations between outputs, this model is suitable for handling compositions. In order to constrain the predictions to sum up to 100 at.%, the predictions are subsequently normalized by division by the sum.

## Data availability

The dataset of this publication is accessible at Zenodo under https://doi.org/10.5281/zenodo.15083643. The code was published on GitLab under https://gitlab.ruhr-uni-bochum.de/mdi/xps-pred.

## Author contributions

F. Thelen: conceptualization, formal analysis, investigation, software, validation, visualization, writing – original draft, writing – review & editing. F. Lourens: conceptualization, sample fabrication, measurement, data curation, formal analysis, writing – original draft, writing – review & editing. A. Ludwig: supervision, resources, writing – original draft, writing – review & editing, project administration.

## Conflict of interest

There are no conflicts to declare.

## Acknowledgements

This work was partially supported by different projects. A. Ludwig and F. Thelen acknowledge funding from the European Union through the European Research Council (ERC) Synergy Grant project 101118768 (DEMI). Views and opinions expressed are however those of the authors only and do not necessarily reflect those of the European Union or ERCEA. Neither the European Union nor the granting authority can be held responsible for them. F. Lourens and A. Ludwig acknowledge funding from DFG SPP 2315 (project number 470309740: LU 1175/36-1).